\documentclass[12pt]{article}
\usepackage{epsf}
\def\be{\begin{equation}}
\def\ee{\end{equation}}
\def\bea{\begin{eqnarray}}
\def\eea{\end{eqnarray}}

\begin{document}
\begin{titlepage}
\begin{center}
{\Large \bf William I. Fine Theoretical Physics Institute \\
University of Minnesota \\}
\end{center}
\vspace{0.2in}
\begin{flushright}
FTPI-MINN-19/19 \\
UMN-TH-3828/19 \\
July 2019 \\
\end{flushright}
\vspace{0.3in}
\begin{center}
{\Large \bf Some decay properties of hidden-charm pentaquarks as baryon-meson molecules 
\\}
\vspace{0.2in}
{\bf  M.B. Voloshin  \\ }
William I. Fine Theoretical Physics Institute, University of
Minnesota,\\ Minneapolis, MN 55455, USA \\
School of Physics and Astronomy, University of Minnesota, Minneapolis, MN 55455, USA \\ and \\
Institute of Theoretical and Experimental Physics, Moscow, 117218, Russia
\\[0.2in]

\end{center}

\vspace{0.2in}

\begin{abstract}
It is pointed out that the previously  suggested interpretation of the hidden-charm pentaquarks $P_c(4312)$, $P_c(4440)$ and $P_c(4457)$ as molecular bound states of $\Sigma_c \bar D$ and $\Sigma_c \bar D^*$ baryon-meson pairs gives rise to specific relations for their decays. In particular, the heavy quark spin symmetry predicts ratios of rates of decays of each of the molecules to $J/\psi p$ and to $\eta_c p$ as well as of the decays to $\Lambda_c \bar D$ and  to $\Lambda_c \bar D^*$. Experimental studies of these relations would thus provide an indicative probe of the molecular structure.
  \end{abstract}
\end{titlepage}

The initial experimental evidence~\cite{lhcbp1,lhcbp2}  for hidden-charm pentaquarks has been recently refined into an observation~\cite{lhcbp3} of three relatively narrow resonances  $P_c(4312)$, $P_c(4440)$ and $P_c(4457)$ in the $J/\psi p$ system produced in the decays $\Lambda_b \to J/\psi p K$. The measured positions of the resonance peaks are just few MeV below the thresholds for hidden-charm baryon-meson channels $\Sigma_c \bar D$ and $\Sigma_c \bar D^*$, and the experimental report~\cite{lhcbp3} strongly suggests an interpretation of the lowest $P_c(4312)$ resonance as a shallow bound $S$ wave state of  $\Sigma_c \bar D$ with the spin-parity $J^P=1/2^-$ and the two higher peaks as similar $S$ wave bound systems of $\Sigma_c \bar D^*$ with $J^P=1/2^-$ and $J^P=3/2^-$. Existence of molecular states with hidden heavy flavor was suggested long ago~\cite{vo76}, and similar heavy meson-antimeson systems have been observed both with hidden charm [e.g. $X(3872)$~\cite{bellex}] and the bottomonium-like $Z_b(10610)$ and $Z_b(10650)$~\cite{bellezb}. (Recent reviews of multiquark hadrons including molecular systems with hidden heavy flavor can be found in Refs.~\cite{dsz,Guo17,Ali17,Liu,mv19}.) It would thus be of a great interest to further test the internal dynamics of the newly found exotic baryons and confirm or rule out their molecular nature. 

In the molecular picture applied to pentaquarks the constituent charmed baryon and (anti)charmed meson move at  distances longer than the size of each, so that they both largely retain their own structure. Thus in the dominant part of the wave function the dynamical correlations between the (anti)quarks are essentially the same as within the individual hadrons, while the interaction between the hadrons resulting in a shallow bound state is of a secondary importance. In the present paper are discussed the consequences of this picture for decays of the pentaquarks, in particular the implications for decays into $\Lambda_c \pi \bar D^{(*)}$, $\Lambda_c \bar D^{(*)}$ and relations between the decays into $J/\psi p$ and $\eta_c p$. The treatment of the latter two types of the decay relies on the Heavy Quark Spin Symmetry (HQSS) applied to the charmed quarks and the resulting predictions are very specific to the assumption that the spins of the charmed quark and the antiquark are correlated with the spins of the light quarks in the baryon and the anti-charmed meson rather than with each other. For this reason the heavy $c \bar c$ pair is in a mixed state with respect to its total spin~\footnote{This behavior is quite similar to that in the bottomoniumlike molecular resonances $Z_b(10610)$ and $Z_b(10650)$~\cite{bellezb} made from bottom meson-antimeson pairs, where due to a mixed spin state of the $b \bar b$ pair~\cite{bgmmv} the resonances decay with comparable rate to final states with ortho- and para- bottomonium.} that has definite projections on the spin states of this pair in the decay channels $J/\psi p$, $\eta_c p$ and also $\Lambda_c \bar D^{(*)}$ that are considered here. Clearly, if the pentaquarks are not (dominantly) molecules, but rather contain a compact $c \bar c$ pair, such as in the suggested baryo-charmonium model~\cite{epp}, the expected spin state of the heavy quark-antiquark pair is completely different, and the relations based on the molecular model would be strongly invalidated. In particular, in the baryo-charmonium model in the limit of HQSS the spin of the compact $c \bar c$ pair is fixed at either one or zero, and a specific pentaquark would decay respectively to either $J/\pi p$ or $\eta_c p$ but into both channels. On the contrary, in the molecular picture discussed here, both decays are possible for one and the same pentaquark, with a specific relation between the rates. In what follows I first discuss the implications of the HQSS for these decays and for the decays to $\Lambda_c \bar D^{(*)}$ and then make brief remarks on the rates of the decays to $\Lambda_c \pi \bar D^{(*)}$ resulting from the underlying known process $\Sigma_c \to \Lambda_c \pi$.

The light-heavy spin structure of the $\Sigma_c$ baryon can be described in terms of the spins of its constituents as
\be
\Sigma^\alpha = (\vec \sigma \cdot \vec \phi)^\alpha_\beta c^\beta~,
\label{ssa}
\ee
where $c^\beta$ is a (nonrelativistic) spinor for the charmed quark, $\vec \phi$ is the polarization amplitude for the spin one light diquark within $\Sigma_c$, $\vec \sigma$ is the vector of Pauli matrices, and Greek indices $\alpha, \beta, \ldots$ are for the spin components. Similarly, the $\bar D$ and $\bar D^*$ spin structure is described as
\be
\bar D = (\bar c \, q)\, ~~~~{\rm and}~~~~  {\bar D^*}_i = (\bar c \, \sigma_i  \, q)
\label{dsa}
\ee
with $q^\alpha$ and $\bar c_\alpha$ standing for the spinors of the light quark $q$ and the $\bar c$ in the heavy anti-meson.  

The spin structure of a widely separated baryon-meson pair is given by the product of the individual spin functions (\ref{ssa}) and (\ref{dsa}). This product contains the matrix $C^\alpha_\beta = c^\alpha \bar c_\beta$ describing the (mixed) spin state of the heavy quark-antiquark pair, and one can expect that this structure applies to a shallow bound state. In the transitions to $J/\psi p$ and $\eta_c p$ the charmonium states are degenerate if the spin-spin interaction of heavy quarks is neglected, i.e. in the HQSS limit, and the spin state of the final heavy quark pair can be described in terms of the amplitudes $\eta$ and $\vec \psi$ for respectively the $\eta_c$ and $J/\psi$ charmonium as
\be
c^\alpha \bar c_\beta = \eta \, \delta^\alpha_\beta + (\vec \sigma \cdot \vec \psi)^\alpha_\beta
\label{csa}
\ee
In the limit of exact HQSS the heavy quark spin is also preserved in the transition, so that one finds relations between the amplitudes by projecting the initial spin state described by $C^\alpha_\beta$ on the charmonium spin state in Eq.(\ref{csa}). One remark is in order regarding the spin state of the light quarks described in this calculation by six components of the product $\phi_i q^\alpha$ generally corresponding to two spin components of a spin 1/2 proton and to four components of the spin 3/2 $\Delta$ resonance, if one limits the consideration of the final light quark hadronis state to a single baryon. The explicit decomposition into the states of definite light baryon spin has the form 
\be
\phi_i q = {1 \over 3} \, \sigma_i \, (\vec \sigma \cdot \vec \phi) \, q + \left [ \phi_i q - {1 \over 3} \, \sigma_i \, (\vec \sigma \cdot \vec \phi) \, q \right ]~,
\label{psa}
\ee
where the first term on the rhs corresponds to the total spin 1/2 while the expression in the straight braces is pure spin 3/2. As is well known, the two spin states of three $u$ and $d$ quarks inside a baryon also have definite isotopic spin. The isospin violation in the molecules can be significant~\cite{gjms} due to the mass differences between the charged and neutral charmed mesons being comparable with the binding energy. Thus the flavor composition of the three-quark state in Eq.(\ref{psa}) is important in a calculation of the relative rate of decays of the pentaquarks with the $\Delta$ resonance and the nucleon in the final state. However any light-flavor structure emerging from the isopin considerations enters as a common factor in the decays with just a proton: $P_c \to J/\psi p$ and $P_c \to \eta_c p$, which factor does not affect the ratio of the rates of these decays. For this reason the isotopic composition of the three light-quark state is suppressed in the present calculation and one can replace the product $\phi_i q$ with just the first term in Eq.(\ref{psa}). Namely, in what follows the substitution is made
\be
\phi_i q \to \vec \sigma_i p~,
\label{pqp}
\ee
where $p^\alpha$ is the spinor for the proton, up to a coefficient that cancels in the ratio considered here.

Using the expressions (\ref{ssa}), (\ref{dsa}), (\ref{csa}) and (\ref{pqp}), the spin structure of the amplitudes for the decay of the lowest pentaquark $P_c(4312)$ considered as a $J^P=1/2^-$  $\Sigma_c \bar D$ molecule: 
\be
P_c^\alpha = \Sigma^\alpha \bar D~,
\label{pca}
\ee  
can thus be written as
\be
A \left [ P_c(4312) \to (c \bar c) p \right ] = \left \{ p^\dagger \sigma_i \left [\eta+ (\vec \sigma  \cdot \vec \psi) \right ]  \sigma_i \, P_c \right \} = 3 (p^\dagger P_c) \eta - \left [ p^\dagger (\vec \sigma  \cdot \vec \psi) P_c \right ]~.
\label{asd}
\ee
The ratio of the decay rates is thus readily found as
\be
{\Gamma[P_c(4312) \to \eta_c p] \over \Gamma[P_c(4312) \to J/\psi p]} = 3~.
\label{rsd}
\ee
In the latter relation the kinematical difference between the two final channels is neglected. The difference in the $S$-wave phase space amounts to approximately 20\% and is due to HQSS symmetry breaking in the masses of $J/\psi$ and $\eta_c$. It is likely that other effects of symmetry breaking amount to similar corrections, so that the theoretical error in Eq.(\ref{rsd}) can be estimated as few tens percent.

An $S$-wave $\Sigma_c \bar D^*$ system can be in a $1/2^-$ or $3/2^-$ state. At present it is not known which of these should be assigned to $P_c(4440)$ and which to $P_c(4457)$ in the molecular picture, since the sign of the spin-spin interaction between the charmed baryon and the (anti)meson is not known. Thus no particular assignment is assumed here and the two higher states will be denoted as $P_{c1}$ and $P_{c3}$ corresponding to the spin 1/2 and 3/2. The former of these states can be written in terms of the spin variables of the baryon and the meson as
\be
P_{c1} =  ( \sigma_i \bar D^*_i) \Sigma~.
\label{pc1}
\ee
Using again the same expressions as in the derivation of Eq.(\ref{asd}), one readily arrives at the relation
\be
A \left[ P_{c1} \to (c \bar c) p \right ] = \left \{ p^\dagger \sigma_i \sigma_j \left [\eta+ (\vec \sigma  \cdot \vec \psi) \right ]  \sigma_i  \sigma_j \, P_{c1} \right \} = -3 (p^\dagger P_c) \eta + 5 \left [ p^\dagger (\vec \sigma  \cdot \vec \psi) P_c \right ]~,
\label{ap1}
\ee 
and finds the ratio of the rates as
\be
{\Gamma[P_{c1} \to \eta_c p] \over \Gamma[P_{c1} \to J/\psi p]} = {3 \over 25}~.
\label{rp1}
\ee
It is quite clear that the spin 3/2 pentaquark cannot decay to $\eta_c p$ in the $S$ wave. Thus any presence of the decay $P_{c3} \to \eta_c p$ would provide a measure  of a $D$ wave dynamics. It should be noted that the predictions (\ref{rsd}) and (\ref{rp1}) each relate the decays of the same pentaquark, and the discussed approach cannot be used in its present form for relating decay rates of different molecular states, both due to apparently different binding energy, resulting in a somewhat different wave functions of the motion, and due to possible difference in the amount of  the previously mentioned  isospin violation between the pentaquarks.

\begin{figure}[ht]
\begin{center}
 \leavevmode
    \epsfxsize=14cm
    \epsfbox{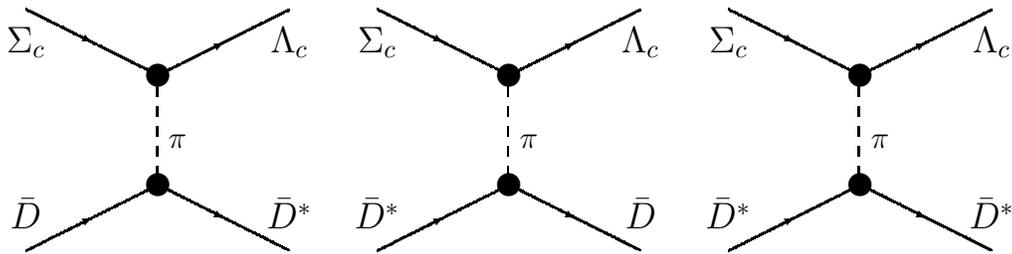}
    \caption{The scattering $\Sigma_c \bar D^{(*)} \to \Lambda_c \bar D^{(*)}$ due to one pion exchange.}
\end{center}
\end{figure} 

Another application of HQSS is to the decays of the molecular pentaquarks to the final states $\Lambda_c \bar D$ and $\Lambda_c \bar D^*$. These decays are allowed by all the conservation laws. Moreover, they are definitely contributed by a one pion exchange as shown in Fig.~1. Both vertices in these graphs are known from the decays $\Sigma_c \to \Lambda_c \pi$ and from $D^* \to D \pi$. One can immediately notice that due to the absence of a $DD\pi$ vertex  there is no contribution of this mechanism to the decay of $P_c(4312) \to \Lambda_c \bar D$, only the decay $P_c(4312) \to \Lambda_c \bar D^*$ is possible. Furthermore, the HQSS relation between the $D^*D \pi$ and $D^* D^* \pi$ vertices gives rise to a relation between the $S$-wave amplitudes of the decays $P_{c1} \to \Lambda_c \bar D$ and $P_{c1} \to \Lambda_c \bar D^*$, while for the spin 3/2 pentaquark $P_{c3}$ only the decay to $\Lambda_c \bar D^*$ is possible in the $S$ wave, and any activity in the $\Lambda_c \bar D$ channel at this resonance would indicate presence of a $D$ wave.

It can be noticed however that the conclusions about absence of the $S$ wave decay $P_c(4312) \to \Lambda_c \bar D$ and relation between the amplitudes of the decays $P_{c1} \to \Lambda_c \bar D$ and $P_{c1} \to \Lambda_c \bar D^*$ are more general than the one pion exchange mechanism and are in fact valid beyond this mechanism as a result of HQSS alone. Indeed, the spin wave function of $\Lambda_c$ has the form 
\be
\Lambda^\alpha  =  \phi_0 c^\alpha~,
\label{lsa}
\ee 
where $\phi_0$ describes the scalar $ud$ diquark. In a soft scattering   $\Sigma_c \to \Lambda_c$ the spin of the charmed quark is preserved according to HQSS, so that an $S$-wave amplitude should be proportional to the expression $(\Lambda^\dagger \sigma_i \Sigma)$ [c.f. Eq.(\ref{ssa})]. In the processes $\Sigma_c \bar D^{(*)} \to \Lambda_c  \bar D^{(*)}$ the recoiling against the charmed baryon state is the $\bar D^{(*)} \to  \bar D^{(*)}$ transition. If one writes in the standard way the quark-antiquark spin matrix for the $\bar D^{(*)}$ mesons as 
\be
q^\alpha \bar c_\beta \sim \bar {\cal D}^\alpha_\beta = \bar D \delta^\alpha_\beta+ (\vec \sigma_i \bar D^*_i)^\alpha_\beta~,
\label{dmx}
\ee
and takes into account that the spin of the heavy quark (in the $\bar D^{(*)}$ meson) is conserved in the scattering, one concludes that the only possible compatible with HQSS expression for the scattering in the $S$ wave has the form
\be
A[\Sigma_c \bar D^{(*)} \to \Lambda_c  \bar D^{(*)}] \propto (\Lambda^\dagger \sigma_i \Sigma) \,  {\rm Tr} \bar {\cal D} \bar {\cal D}^\dagger \sigma_i \propto (\Lambda^\dagger \sigma_i \Sigma)  \left [\bar D^\dagger \bar D^*_i + (\bar D^*_i)^\dagger \bar D - {\rm i} \epsilon_{ijk} (\bar D^*_j)^\dagger \bar D^*_k \right ]\,. 
\label{asc}
\ee
One can thus readily notice that the $\bar D^{(*)}$ (axial) vector factor in this expression is exactly the same as entering the pion emission vertex by the  $\bar D^{(*)}$ mesons.
This expression contains no $\bar D \to \bar D$ transitions, so that the decay $P_c(4312) \to \Lambda_c D$ is forbidden by HQSS. The relation between the rates of the $S$ wave decays of the spin 1/2 molecular pentaquark $P_{c1}$ can also be directly found as
\be
{\Gamma(P_{c1} \to \Lambda_c \bar D^*) \over \Gamma(P_{c1} \to \Lambda_c \bar D)} ={ 4 \over 3}~,
\label{rp1l}
\ee
modulo  effects of HQSS breaking. In this case however the kinematical differences can be somewhat more significant than for the case of Eq.(\ref{rsd}),  amounting to about 40\% suppression of the ratio (\ref{rp1l}) if $P_{c1}$ is identified as $P_c(4440)$. 

The absolute rate of the decays of molecular pentaquarks to $\Lambda_c \bar D^{(*)}$ is currently quite uncertain. An estimate based on the pion exchange shown in Fig.~1 depends on the unknown short-distance behavior of the wave function of the molecules and its reliability is unclear.  This issue may be somewhat similar to the known example of the bottomoniumlike $B^* \bar B^*$ molecular resonance $Z_b(10650)$ , where a straightforward estimate based on one pion exchange gives the rate of the decay $Z_b(10650) \to B^* \bar B +$c.c. about an order of magnitude larger than the experimental upper limit~\cite{bellezbb}. The reason for such suppression is not known and the behavior is described   in terms of a form factor suppression~\cite{mv16} of the pion exchange or as an effect of a contact term~\cite{wangetal} effectively canceling the pion contribution. Given that the puzzle of that suppression is not yet resolved, it is currently not clear if a similar behavior may take place in the decays of hidden-charm molecules to $\Lambda_c \bar D^{(*)}$. As a result it appears that existence of these open decay channels with a significant energy release does not necessarily preclude the interpretation of the observed relatively narrow pentaquarks as $\Sigma_c \bar D^{(*)}$ molecules.

One more probe of possible molecular structure of the discussed pentaquarks is provided by the existence of the decay $\Sigma_c \to \Lambda_c \pi$ with the width of about 2\,MeV that for molecular states should result in decays $P_c \to \Lambda_c \pi \bar D^{(*)}$. The binding energy in the pentaquarks $P_c(4312)$ and $P_c(4457)$ is somewhat smaller than the energy released in $\Sigma_c \to \Lambda_c \pi$:  $M(\Sigma_c) - M(\Lambda_c) - m(\pi) \approx 30\,$MeV. It can thus be expected that in these molecular states the decays of $\Sigma_c$ are only slightly affected by the binding, so that the width of the decays $P_c(4312) \to \Lambda_c \pi \bar D$ and $P_c(4457) \to \Lambda_c \pi \bar D^*$ should be about the same as for a free $\Sigma_c$ and thus make a significant fraction of the measured~\cite{lhcbp3} total widths of these pentaquark resonances, respectively $(9.8 \pm 2.7^{+3.7}_{-4.5})\,$MeV and $(6.4 \pm 2.0^{+5.7}_{-1.9})\,$MeV. The expected rate of similar decay for the $P_c(4440)$ resonance should apparently be suppressed, since its lower mass (stronger binding in the molecular picture) leaves only small phase space for the final state $\Lambda_c \pi \bar D^*$. It should be also noted that the decays with the `wron spin' heavy meson, such as $P_c(4440) \to \Lambda_c \pi D$ or $P_c(4457) \to \Lambda_c \pi D$, are not directly related to the decays of a loosely bound $\Sigma_c$ and cannot be treated in the simplistic manner as discussed here.

The main points of this paper are summarized as follows. In a shallow bound molecule the spin correlations are preserved as in the constituent hadrons. Thus the heavy quark-antiquark  pair is in a mixed spin state and $c \bar c$ states with spin 0 as well as spin 1 can be produced in the decays of molecular pentaquarks to charmonium plus a nucleon. Moreover, in the HQSS limit the lowest charmonium states $\eta_c$ and $J/\psi$ are degenerate, so that their relative yield in such decays is fixed by the quantum numbers of the bound molecule. In this way the ratio of the yield in decays of the $P_c(4312)$ viewed as an $S$-wave $\Sigma_c \bar D$ system is given by Eq.(\ref{rsd}) and for a spin 1/2 $\Sigma_c \bar D^*$ molecule the relation is in Eq.(\ref{rp1}). Both relations are expected to have accuracy of the HQSS limit for charmed quark, i.e. likely few tens percent. It is essential that there should be comparable yield in both decay channels. This is in contrast with the expected behavior in e.g. the baryo-charmonium model~\cite{epp}, where it is either ortho- or para- charmonium embedded in a light hadron which thus is dominantly produced in the decay. Furthermore, the spin correlations in an $S$-wave molecule and the HQSS result in definite predictions for relative amplitudes of decays to $\Lambda_c \bar D$ and $\Lambda_c \bar D^*$, namely, the former decay is forbidden for the lowest $P_c(4312)$ state and the relative yield in the two channels in the decays of the spin 1/2 $\Sigma_c \bar D^*$ state is given by Eq.(\ref{rp1l}). Any activity in the channels forbidden for $S$-wave systems, e.g. the decays of the spin 3/2 pentaquark to either $\eta_c p$ or $\Lambda_c \bar D$ would  provide an access to the $D$-wave motion in the considered systems. The notion of a loosely bound state also implies that the $\Sigma_c$ hyperon  can undergo its usual strong decay $\Sigma_c \to \Lambda_c \pi$ with the rate approximately equal to that of the free particle decay inasmuch as the binding energy is small in comparison with the available energy. The latter appears to be the case for $P_c(4312)$ and $P_c(4457)$, but not for the stronger bound $P_c(4440)$. Thus one can expect the decays $P_c(4312) \to \Lambda_c \pi \bar D$ and $P_c(4440) \to \Lambda_c \bar D^*$ to make a considerable fraction of the total widths of the resonances. Experimental studies of the discussed here decay properties of the hidden-charm pentaquarks can provide telltale signs confirming or rejecting their molecular nature.

I thank A.~Bondar for illuminating discussions. This work is supported in part by U.S. Department of Energy Grant No.\ DE-SC0011842.

\end{document}